\documentclass [letterpaper]{IEEEtran}
\usepackage[dvipsnames]{xcolor}
\usepackage{graphicx,
	psfrag,
	epsfig,
	epstopdf,
	amsthm,
        amsmath,
	amssymb,
	url,
	subcaption,
	caption,
	balance,
	enumerate,
	setspace,
	tikz,
	pgfplots
}
\usepackage{amsmath}
\usepackage{algpseudocode}
\usepackage[nospace,noadjust]{cite}
\usepackage{pbox}
\usepackage{multirow}
\usepackage{adjustbox}
\usepackage{array}
\usepackage{comment}
\usepackage{booktabs}
\usepackage{float}
\usepackage{tabularx}
\usepackage{breqn}
\usepackage{bbm}
\usepackage{nomencl}
\usepackage[ruled,vlined]{algorithm2e} 
\usepackage{amsmath}                  
\usepackage{amssymb}                  
\usepackage{amsthm}                   

\makenomenclature
\usetikzlibrary{shapes.geometric}

\newcommand{\cref}[1]{Constraint~\ref{#1}}

\newcommand{\ignore}[1]{}
\usepackage[paperwidth=8.5in, paperheight=11in,top=1.1in, bottom=0.99in, left=0.575in, right=0.6in]{geometry}

\begin{document}

\title{Raft Distributed System for Multi-access Edge Computing Sharing Resources}

\author{
	\IEEEauthorblockN{Zain Khaliq\IEEEauthorrefmark{1}, Ahmed Refaey Hussein \IEEEauthorrefmark{2}}\\

	\IEEEauthorblockA{\IEEEauthorrefmark{1} Manhattan College, New York, USA.}\\
	\IEEEauthorblockA{\IEEEauthorrefmark{2} University of Guelph, Ontario, Canada.}}

\maketitle

\maketitle
\begin{abstract}

Researchers all over the world are employing a variety of analysis approaches in attempt to provide a safer and faster solution for sharing resources via a Multi-access Edge Computing system. Multi-access Edge Computing (MEC) is a job-sharing method within the edge server network whose main aim is to maximize the pace of the computing process, resulting in a more powerful and enhanced user experience. Although there are many other options when it comes to determining the fastest method for computing processes, our paper introduces a rather more extensive change to the system model to assure no data loss and/or task failure due to any scrutiny in the edge node cluster. RAFT, a powerful consensus algorithm, can be used to introduce an auction theory approach in our system, which enables the edge device to make the best decision possible regarding how to respond to a request from the client. Through the use of the RAFT consensus, blockchain may be used to improve the safety, security, and efficiency of applications by deploying it on trustful edge base stations. In addition to discussing the best-distributed system approach for our (MEC) system, a Deep Deterministic Policy Gradient (DDPG) algorithm is also presented in order to reduce overall system latency. Assumed in our proposal is the existence of a cluster of N Edge nodes, each containing a series of tasks that require execution. A DDPG algorithm is implemented in this cluster so that an auction can be held within the cluster of edge nodes to decide which edge node is best suited for performing the task provided by the client.  

\end{abstract}


\section{Introduction}

Distributed systems are computer systems in which numerous components on various computers communicate with one another to form a single coherent system. Distributed systems are often implemented due to there horizontal scalability, reliability and performance. New computing and telecommunications technologies based on high-speed networking have opened up new opportunities for complex application development \cite{9} \cite{b}. However, often times complex architectural design of these distributed system can effect the scheduling, latency and/or transparency of the task that is being executed on these networks. Some of these complicated distributed systems applications are used in our daily lives, and we virtually never think about the extensive engineering that goes into making these applications so powerful. Mobile Communication Networks \cite{10}\cite{f}, Online Banking \cite{11}, Online Gaming Servers \cite{12}, and a wide range of other applications are just a few examples of distributed systems in use. In a distributed system, there are many software applications that are distributed across multiple nodes. We build a distributed consensus algorithm where a collection of nodes work as a coherent group and be able to still execute all tasks even if some members in the system are inactive or fail to respond. Some of the many consensus algorithms, Paxos algorithm is the most popular distributed consensus algorithm. However, due to its complexity, it is very hard to implement this algorithm in modern software solutions. Raft consensus algorithm is a more modern \cite{g}\cite{h}, reliable, and relatively less complicated distributed consensus algorithm that is based on a leader-based driven consensus model. Here a distinguished leader is elected in the system that takes full responsibility of managing the distributed system cluster \cite{d}\cite{e}.  

To better understand the concept of a consensus algorithm we introduce blockchain. A blockchain network is a distributed database or ledger maintained by a group of independent users that do not trust each other \cite{4}. Bitcoin is one example of this network where it became the first currency system that is entirely decentralized and independent of any monetary authority \cite{13}. In this network, the use of cryptography makes all transactions among one another without the oversight of a centralized power \cite{4}. Every valid transaction is recorded in whats known as a ledger, and each user receives a copy \cite{4}\cite{i}. Consensus is the process by which a transaction is accepted as valid and recorded in the ledger. When transactions are verified, they are added to the ledger in a block-by-block format. The chronological order of each valid and approved transaction is whats known as the blockchain, in which each node is connected to the next based on its transaction execution. As the number of blocks in this chain grow the computational execution of the consensus algorithm becomes more difficult, which often lead to performance and complexity issues in the system \cite{i}\cite{j}. 

In modern day software solutions, majority of modern day systems and web applications are hosted in the cloud. Though this has given us a great roadmap ahead, we face constraints regarding the cloud as well, which is why edge computing has been introduced in the past few years. Edge computing is a system that solves the latency issue between the client and the cloud. The edge server acts as a middleman in being able to fetch data quick from itself, or from the cloud server and back to the client(s). Cloud computing offers platforms that assist in configuring a server or a number of servers in order to host an application, while edge computing consists of servers that are closely linked with the client \cite{k}. 

In this paper we introduce a blockchain based solution for edge computing, where we introduce the raft consensus algorithm to handle various requests made by the client. Specifically, to solve performance and complexity issues in the system in this protocol, we introduce the Deep Deterministic Policy Gradient (DDPG) learning algorithm for our distributed system. Rather than depicting a random election like the conventional raft distributed system, our DDPG-based Multi-access Edge Computing Sharing Resource training algorithm improves latency in our system model \cite{k}\cite{l}\cite{m}. Through its introduction as an optimization problem, DDPG assists the raft protocol by allowing it to elect the most resourceful node in the cluster as its leader to handle various requests made by the client. The Bellman equation and off-policy data are used by DDPG to learn the Q-function, which is then used to learn the policy of raft consensus architecture, hence improving performance and complexity issues in the overall system \cite{mm} \cite{n}.

\section{Related Work}
A collection of IEEE papers have been studied in order to properly understand the Raft Algorithm. The problem of trying to find the best consensus algorithm isn't a hot new topic. In fact solving for consensus dates all the way back to 1988, which was demonstrated by Lynch, Dwork and Stockmeyer \cite{3} \cite{o} \cite{nn}. They introduced the "partial synchrony" in a distributed system, in which they use new protocols for a fault tolerant "distributed clocks" where common notion of time amongst partially synchronous processors is reached \cite{3} \cite{oo}. Later in 1989, paxos protocol was submitted, which got its name from a fictional legislative consensus system used on the Paxos island in Greece. Paxos algorithm is a protocol to solving consensus in a network of unreliable or faulty processors. When Paxos was first introduced, Lamport's metaphoric language explaining the algorithm made it very difficult to understand. Lamport later wrote another version of Paxos and called it "Paxos Made Simple". However, "Paxos Made Simple" turned out to be not so simple to understand. After many attempts to implement the Paxos algorithm in practical use case, researchers came to discover the raft algorithm. Raft is a consensus algorithm that was developed to understand the Paxos algorithm and for its easy deployment. Similar to Paxos in performance and fault-tolerance, raft differs that it tends to solve independent sub-problems and while advocating its use case for practical systems \cite{2} \cite{p}.

When comparing the conventional, centralized distributed solution, blockchain by far overtakes its competition in various ways such as: immutability, enhanced security, fault tolerance, and its transparency \cite{4} \cite{pp}. Although a decentralized network has its perks, the nature of its use case of blockchain puts a huge constraint on its performance level. One of the prime examples of a lack in performance is Bitcoin. With a transaction time that takes roughly 10 minutes to be completed, Bitcoin has a low throughput of 7 transactions per second (TPS) \cite{6}. With per \cite{l}formance being the biggest concern many new consensus algorithms are being designed to solve this issue. Numerous research are investigating ways to make blockchain-based IoT applications run faster. Most of them employ proof-of-work (PoW) consensus and MEC offloading to minimize burden \cite{7} \cite{q}. However, much resources are lost in PoW consensus in order to combat rogue nodes \cite{8}.

\section{Raft Algorithm}

Raft is a consensus mechanism for keeping track of a replicated ledger across all nodes \cite{1}. Similar to the Paxos algorithm in the fault-tolerance and performance scenario, Raft algorithm works in a similar way, however, Raft algorithm is much simple in terms of its implementation when compared to the Paxos algorithm. To achieve consensus multiple servers must agree on the same information,assuming the consensus protocol holds the following four features: Validity, Agreement, Termination and Integrity.\\
Raft algorithm can be applied to solve both a single server system problem and/or Multiple Server System problem. In this paper we explore the Raft algorithm in a multiple-server system, which tends to be the most used architecture in modern-day systems. A client interacts with two types of numerous servers in a multiple-server system: Symmetric and Asymmetric. Any of the several servers in a Symmetric system can react to the client, whereas the other servers sync with the server that's also responding to the client's request. In an asymmetric system, only the elected server may reply to the client, while the other servers in the node are in sync with it.\\
There are three significant states of server nodes that define the Raft algorithm \cite{1}: Leader, Follower, and the Candidate. There is exactly one leader, while the other servers are considered followers in a normal operation \cite{2}. As the leader handles all client requests, the followers issue no request on their own, but respond to the requests made by the leader and the candidates \cite{2}. Candidates state is used to elect new leaders. Raft divides time into terms of arbitrary length to maintain the server status. The term number is passed around and maintained by every node. These terms are numbered with consecutive integers, where each term begins with an election to choose a new leader \cite{2}. During the election, the candidate asks for the vote and if a majority is gathered the candidate becomes the leader, otherwise, the term ends with no leader, known as split vote. Raft algorithm also uses two types of Remote Procedure Calls (RPCs) to carry out its communication function: RequestVotes and AppendEntries. RequestVotes RPC is sent by the Candidate nodes to gather votes during an election, while AppendEntries is used by the Leader node for create the log entries. AppendEntries RPC is also used as a heartbeat mechanism to check if a server is still up. 

\subsection{Leader election}

If the current leader fails or the raft algorithm is initially initialized, elections are held in the raft. When a new term in the cluster begins, it is for an indefinite amount of time on the server where the leader is chosen. A "leader election" kicks off each term. If the election is successful, the term will continue as usual; if the election is unsuccessful, a new term will begin with a fresh election. 

The candidate server starts the leader election, and the server only becomes a candidate if the leader does not communicate with it during the election timeout. During the election timeout, the candidate assumes there is no leader and conducts its own election by boosting the term counter and casting itself as the new leader, while simultaneously sending a message to all other servers soliciting votes. There are three main cases that decides the winner of the election. Since the election is based on first-come-first-served basis, only one vote per term will be cast by the server. If the candidate receives a message from another server with a term number greater than the candidate's, the election is lost, and the candidate becomes a follower. However, candidate will be recognized as a new leader if it receives majority of the votes. If neither of the above two cases occurs this is know as a "split vote", where a new election will begin with a new term.

Often time the Raft algorithm uses the randomized election timeout to solve the split vote issues. Since the the timer on each server will be different, the first server will time out first will become the leader and send a heartbeat message to other server before any of the followers become the candidates.

\subsection{Log Replication}

Once a leader has been elected \cite{2},  the server starts listening to the clients requests. Here, every client call is submitted consisting of a command that the replicated state machines run.. Each time a new entry is added to the log of the leader, the request is passed on to the follower as a "AppendEntries" messages. If the follower is unavailable the leader will be sending the "AppendEntries" notifications until all of the followers have saved all of the log entries.  

After all the entries have been replicated by the followers, and the leader receives a confirmation, the leader's local machine is then accessed, and is asked to be deemed committed.\cite{2}. Shortly after the leader commits the log entry, all the followers apply the entry to its local machine, thus insuring the logs from all the hosts in the cluster are replicated. 

If the leader is to crash for any reason, a new leader will then handle the logs that are left inconsistent by the previous leader by requiring followers to record their individual log. To restore the log consistency in a failed cluster, the leader compares its log with the log from its followers and finds the last entry on which they had agreed. It then deletes all entries in the follower log that come after this crucial entry and replaces them with its own log entries.

\section{System Model}

We assume in the proposed system architecture (Fig coming soon) there are $N$ cluster of Edge node, each of which maintains series of tasks that are required to be executed and can be indexed by $n = 1,2,….,N$. It is also assumed that $N$ cluster of Edge node are optimized on a Hyperledger Fabric network, where the main source of requests is coming from the cloud level. The cloud level of this system architect consists of a system that has the ability to shortlist some edge servers, in order to assign the tasks to the most capable performing edge server. 
\begin{table}[ht!]
\centering
\small
\caption{Symbols of Equations}
\begin{tabular}{|p{1.5cm}|p{6.5cm}|}
 \hline
 \multicolumn{2}{|c|}{\textbf{Parameters}} \\
 \hline
 $\tau^{\mathrm{TML}}$          & Transactions migration latency \\
 $\left|Z_{l}(\tau)\right|$     & The number of items in the set \\
 $K$                            & Transaction's size \\
 $\beta$                        & Pace of transmission of a fiber connection between two leaders \\
 $\tau^{\mathrm{CLBG}}$         & Expected computing latency of block generation \\
 $N$                            & Number of transactions \\
 $\nu$                          & Average CPU cycles required per hash operation \\
 $\iota$                        & Number of (CPU) cycles per unit time that $l$ possesses \\
 $l$                            & Current leader in the Raft protocol \\
 $\theta$                       & Parameters from actor network \\
 $f$                            & Follower \\
 $l$                            & Leader \\
 $R$                            & Rate \\
 $R_{l, f}$                     & Wireless link's transmission rate between Leader $l$ and Follower $f$ \\
 $F_{l}$                        & Number of followers where $f$ is 1 follower \\
 $B$                            & Total bandwidth \\
 $\rho_{l, f}$                  & SNR of the wireless connection between transmitter $l$ and receiver $f$ \\
 $W$                            & Total transmission power divided for $F_{l}$ followers \\
 $M_{0}$                        & White noise's average power spectral density \\
 $d_{l, f}$                     & Distance between Leader $l$ and Follower $f$ \\
 $\varrho$                      & Path loss ratio \\
 $\Upsilon$                     & Log-normal distribution \\
 $\tau_{l}^{\mathrm{RCB}}$      & Consensus on the new block \\
 $H$                            & Size of an AppendEntries message \\
 $U$                            & Size of the confirmation message \\
 $\zeta_{k}$                    & $\left\lfloor V_{k} / 2\right\rfloor$ \\
 $\psi$                         & Parameters from critic network \\
 $\mathcal{G}$                  & Gaussian distribution \\
 $S$                            & Current state of actor \\
 $S_{0}$                        & Initial state \\
 $\delta$                       & Experience deque \\
 $\gamma$                       & Index for experience deque \\
 $\Gamma$                       & Decision epoch \\
 $\pi^{*}$                      & Optimal policy gradient \\
 $A$                            & Action \\
 $g$                            & Noise \\
 $S_{\Gamma+1}$                 & Next state \\
 $R$                            & Reward \\
 $\alpha$                       & Learning rate \\
 $\lambda$                      & Discount factor \\
 $y$                            & New value for critic network \\
 $\nabla \mathcal{L}$           & Gradient for critic network \\
 $\nabla \mathcal{J}$           & Gradient for actor network \\
 \hline
\end{tabular}
\end{table}

The shortlisted edge server hosts a competition among themselves to decide a leader among its cluster to perform the task provided via the cloud level. Raft protocol is used to carry out this election process. After initiating the raft protocol, term leader N from the n-th cluster of the edge servers is selected to execute tasks. 

As term leader $N$ is selected from $n$-th cluster, every edge node excluding the leader node becomes a follower in the $n$-th cluster, thus creating a gateway of communication between all the followers and the leader to commit a copy of the executed task on all nodes in the nth-cluster. We denote this gateway of communication between cluster $n$ as $G_n$.  Let * denote the epoch when a new block is generated where the task execution is recorded and sent to the blockchain to be committed. 

Considering the main objective of this system model is for the edge server to perform task based on its availability, latency however between the communication between the cloud level as well as the gateway communication between other edge nodes can cause some performance inefficiency in the overall system. 

We consider the latency caused by the raft protocol in our Hyperledger network, where we consider three scenarios in our system that cause high latency. Assuming the leader to execute tasks in our Hyperledger network is already elected, the first form of latency is caused via migration. This occurs from the cloud level once raft protocol is complete and the leader is declared. During this time the leader receives execution tasks from the cloud level in order to execute them and generate the result on a new block. The communication between the cloud level and the leader of the edge nodes causes latency which can be computed from the following equation: (ADDD EQ 1 HERE)   

\begin{equation}
\tau_{l}^{\mathrm{TML}}=\frac{\left|Z{l}(\Gamma)\right| K}{\beta}
\end{equation}


Once the task is performed by the leader it needs to generate a new block in the Hyperledger network in order to save the results.  We calculate the expected computing latency of a new block generation from the following equation: (Equation 2) 

\begin{equation}
\tau_{l}^{\mathrm{CLBG}}=\frac{\left(N+2^{N}-1\right) \nu}{\iota_{l}}
\end{equation}


Finally, once the leader generates a new block it needs to reach a consensus with all the other nodes in the cluster, in order to commit the result on the Hyperledger network, this way every edge node in the cluster can keep a copy of the result. The latency of sending the message to all the followers and receiving confirmation back from all the followers can be computed via following two equations: ( EQ 3 and EQ 4 )  

\begin{equation}
R_{l, f}=\frac{B}{F{l}} \log \left(1+\rho{l, f}\right), 1 \leq f \leq F_{l}
\end{equation}


\begin{equation}
\rho{l, f}=\frac{W d{l, f}^{-\varrho} \Upsilon}{B M_{0}}
\end{equation}

Using the metrics from the equations above, we derive a new equation to compute the total block consensus latency: 

\begin{equation}
\tau_{l}^{\mathrm{RCB}}=\frac{H}{R_{l, \zeta_{l}}}+\frac{U}{R_{\zeta_{l}, l}}
\end{equation}

\section{Proposed DDPG Model}

An optimization problem is formed to further improve latency in our system model. 
We consider a deep reinforcement learning approach where a learning agent interacts with its environment to improve its performance based on its own experience in the network. We address this experimentation in our paper through the use of Deep Deterministic Policy Gradient (DDPG), a reinforcement learning technique that combines both Q-learning and Policy gradients. DDPG consists of two models, an “actor” and a “critic”. This technique is often referred to as the actor-critic technique, where the actor is a policy network that takes the state as input and outputs the exact action. In contrast, the critic is a Q-value network that takes a state and action as an input and outputs the Q-value. In summary, the actor is responsible for carrying out the actions in the network, while the critic is responsible for critiquing and correcting/providing feedback based on the previous action(s) performed by the actor in the network. The actor and the critic work together to both train the model through the actor’s actions and correct them based on the critic’s feedback, thus resulting in an improved optimal policy for future actions \cite{qq}. 

\subsection{Training}

Regarding the initial state of the model in this paper, consider the starting state as $\mathcal{S}$, where $S_{0}=\{0,0, \ldots, 0\}$. We also declare our experience deque as $\delta$, which represents an array of previous experiences from which the actor learns and continuously improves the experience of its learning in the network. The starting experience deque in this paper is set to $\delta=\varnothing$, at this point it’s considered that the network has not been part of any previous state or action therefore no previous experience has been attained.  To initiate the experience deque, the actor will start at an initial state  $S_{0}$, which will consider a random action whilst in that state. Thus, after the current state $S{t}$ equates to the future state $S{t+1}$, the $S_{0}$ will be sent to the experience deque. In the algorithm below we show that the number of epochs is measured and indexed using $\gamma$ and the total number of epochs/iterations is $\Gamma_{\max }$. We declare our final output as $\pi^{*}$, which is the optimal policy. The optimal policy represents the best possible action for the actor to perform in our DDPG model.

Once initialized, or while $\Gamma \neq \Gamma_{\max }$, the model performs a routine set of steps until the statement is no longer true. This routine depends on the number of epochs that have been completed until it reaches the maximum number of epochs that is preset (500 epochs in this paper). To further expand on this initial state of this routine, the actor will choose a random action on the Gaussian distribution plot through the following equation:\\
\begin{equation}
\mathbf{A}=\pi(S \mid \psi)+g, where g \sim \mathcal{G}
\end{equation}
Since the goal of the algorithm is to improve the action after an updated optimal policy, we refine the actions using the following equation:\\
\begin{equation}
\mathbf{A}^{*}=\arg \max_{\tilde{\mathbf{A}} \in \mathbb{A}} \hat{Q}(S, \hat{\mathbf{A}}), where \mathbb{A}=\left\{\tilde{\mathbf{A}} \mid\|\tilde{\mathbf{A}}-\pi\|_{2} \leq \delta\right\}
\end{equation}

Upon refining its action for the first time from its initial state $S_{0}$, the algorithm allows every leader to perform a random action $\mathbf{A}$ that has been presented at the start of the routine. Following the action $\mathbf{A}$ that is taken by the leader in its initial state, it is presented with a reward $R$, which the algorithm utilizes to continuously improve its future state $S{t+1}$.\\

When transitioning from a current state model to a future state the following equation is applied:\\
\begin{equation}
S_{\Gamma+1}=f(S, \mathbf{A}) 
\end{equation}
where $\mathcal{S}$ is the current state that the model is in. Here, $\mathbf{A}$ represents the current action performed to arrive at the current state. Additionally, $S_{\Gamma+1}$ is the future state in which you would want to reach. \\

The observation series resumes after the changeover $\left(S, \mathbf{A}, R, S_{\Gamma+1}\right)$ is pushed to the deque of experience $\delta$. Here the experience deque represents a memory of previous occurrences in the neural network, where during the time of its training the algorithm is continuously learning. The algorithm improves based on where it previously was, where it is now, the reward it received due to the action, and the actions that it took in order to receive the reward. Once the transition to the new state is complete, the new state becomes the current state and the new epoch becomes the current epoch as shown in below equations :

\begin{equation}
\Gamma \leftarrow \Gamma+1
\end{equation}
\begin{equation}
S \leftarrow S_{\Gamma+1}
\end{equation}

\subsection{Testing}
Once the experience queue has obtained sufficient experience to initiate testing and reinforce itself based on new actions and new states, a random mini batch $e \sim \delta$ is selected. Here, $\delta$ represents the experience queue and the information that is stored into memory from previous actions taken at the start of the initialization of the model. The experience deque is used to assist the actor in choosing the best action. The next step in the process is obtaining the target action batch by the target actor network via $\mathbf{A}^{\prime}=\pi^{\prime}(S), where S \sim e$. Here, note that $\mathbf{A}$ is our current action and $\mathbf{A'}$ the predicted action. Essentially this equation simulates the result of the action prior to the actor taking the action. Parallel to this, the objective critic networks obtain the objective action value batch through $Q^{\prime}\left(S, \mathbf{A}^{\prime}\right)$. Here critic network proposes the best action for the actor to be taken. Using this information, target value $y$ is computed using the equation $y=R+\lambda Q^{\prime}$. When the reward and the critics feedback is combined, you are presented with $y$, which essentially equals to the expected value after the actor has performed the action. Furthermore, this information can be used to form the optimization problem, $\mathcal{L}(\theta)=\mathbb{E}\left\{(Q-y)^{2}\right\}$. This equation is responsible for displaying the accuracy of the model's prediction and using backpropagation to further improve the model. Based on the loss function that we used to formulate our optimization problem, the gradient $\nabla \mathcal{L}$ of the critic network is calculated. The gradient is responsible for telling the network if it is getting better or worse based on the action that the actor has taken. Here, the gradient needs to be the low as possible, where a small gradient means very minor error, therefor the actor is picking the best action. As the value of the gradient approaches 0, the error in the critic network decreases, hence the actor can accommodate better to its critic network predictions. The parameters for the critic network are updated as follows:\\
\begin{equation}
\theta \leftarrow \theta-\alpha_{1} \nabla \mathcal{L}
\end{equation}
The equation above will update the parameters based on the learning rate and the gradient that have been calculated prior in the algorithm. The learning rate is responsible for the "speed" at which you set for the error to approach the global minima. It is significant to carefully decide the value that is set for the learning rate. For instance, if the learning rate is set too high, this will yield a high probability of "overshooting" the global minima. In contrast, if the learning rate is set too low, it will take a longer time to approach the global minima. 
Upon updating the parameters, it is essential in getting the raw action batch. Simply, the algorithm takes the actions that the actor would take as well as the actions that are predicted from the actor-network and the critic network, respectively. Finally, the gradient of the actor-network is calculated based on the output from the critic network.
\begin{equation}
\nabla J=\frac{\partial Q}{\partial \psi}=\frac{\partial Q}{\partial \mathbf{A}} \frac{\partial \pi}{\partial \psi}
\end{equation}

This output depends on the feedback regarding the actor from the critic's predictions $\partial \mathbf{A}$, therefore its result depends on the change in the policy gradient. Since $\partial \pi$ is dependent on the overall parameter of the actor-network, the network is now optimized. To conclude this optimization problem for our system model, parameters on the actor-network are updated, which in return would cause a "chain reaction" and reduce the overall error in the actor network. This final step is accomplished through the utilization of the Gradient Ascent $\psi \leftarrow \psi+\alpha_{2} \nabla J$, where $\alpha_{2}$ is the actor network's learning rate. In conclusion, the target network's settings are updated accordingly via the following equation:\\
\begin{equation}
\psi^{\prime}=\kappa \psi^{\prime}+(1-\kappa) \psi, \theta^{\prime}=\kappa \theta^{\prime}+(1-\kappa) \theta
\end{equation}

\subsection{DDPG Algorithm}

This algorithm presents a Deep Deterministic Policy Gradient (DDPG)-based resource training framework designed for multi-access edge computing environments. It begins by initializing key components, including parameters, states, and an experience replay buffer, to set the foundation for efficient training. Following this, the algorithm employs a policy network to sample continuous actions, adding exploration noise to encourage diverse decision-making. These actions are then refined using a critic-network, ensuring that only those with the highest expected rewards are selected for execution \cite{a}.

Subsequently, the algorithm interacts with the environment as leaders execute actions, observe corresponding rewards, and transition to new states. These transitions, along with associated rewards, are stored in an experience deque to facilitate learning. During the training phase, the critic network updates its value function by minimizing a loss based on the observed data, while the actor-network refines the policy through policy gradient optimization. To ensure stability and prevent overfitting, the algorithm also updates the target actor and critic networks incrementally, using a weighted approach \cite {a} \cite{c}.

\begin{algorithm}[h]
\small
\SetAlgoLined
\SetKwInOut{Input}{Input}
\SetKwInOut{Output}{Output}

\Input{
    Initialize parameters $\psi, \theta \sim \mathcal{G}$, where $\mathcal{G}$ is a Gaussian distribution. \\
    Set initial state $S = S_{0} = \{0, 0, \ldots, 0\}$. \\
    Initialize experience deque $\delta = \varnothing$. \\
    Set epoch threshold $\gamma$ and maximum iterations $\Gamma_{\max}$. 
}
\Output{Optimal policy $\pi^{*}$}

\BlankLine
\textbf{1. Initialization:} \\
Set iteration counter $\Gamma = 0$.

\BlankLine
\textbf{2. Training Loop:} \\
\While{$\Gamma \neq \Gamma_{\max}$}{
    \Indp
    Sample a continuous action:
    \[
    \mathbf{A} = \pi(S \mid \psi) + g, \quad g \sim \mathcal{G}
    \]

    Update action refinement:
    \[
    \mathbf{A}^{*} = \arg \max_{\tilde{\mathbf{A}} \in \mathbf{A}} \hat{Q}(S, \tilde{\mathbf{A}})
    \]

    Execute action $\mathbf{A}$, observe reward $R$, and transition to the next state:
    \[
    S_{\Gamma+1} = f(S, \mathbf{A})
    \]

    Store the observation $(S, \mathbf{A}, R, S_{\Gamma+1})$ in $\delta$. \\
    Update $\Gamma$ and $S$:
    \[
    \Gamma \gets \Gamma + 1, \quad S \gets S_{\Gamma+1}
    \]
    \Indm

    \BlankLine
    \If{$\Gamma > \gamma$}{
        \Indp
        Sample a mini-batch $e$ from $\delta$. \\
        Compute target action batch:
        \[
        \mathbf{A}^{\prime} = \pi^{\prime}(S), \quad S \sim e
        \]

        Compute target action value:
        \[
        Q^{\prime}(S, \mathbf{A}^{\prime})
        \]

        Compute target value:
        \[
        y = R + \lambda Q^{\prime}
        \]

        Compute critic gradient $\nabla \mathcal{L}$ and update critic network:
        \[
        \theta \gets \theta - \alpha_{1} \nabla \mathcal{L}
        \]

        Compute actor action batch:
        \[
        \mathbf{A} = \pi(S), \quad S \sim e
        \]

        Compute actor gradient:
        \[
        \nabla J = \frac{\partial Q}{\partial \mathbf{A}} \frac{\partial \pi}{\partial \psi}
        \]

        Update actor parameters:
        \[
        \psi \gets \psi + \alpha_{2} \nabla J
        \]

        Update target networks:
        \[
        \psi^{\prime} \gets \kappa \psi^{\prime} + (1 - \kappa) \psi
        \]
        \[
        \theta^{\prime} \gets \kappa \theta^{\prime} + (1 - \kappa) \theta
        \]
        \Indm
    }
}
\caption{DDPG-based Multi-access Edge Computing Resource Training with Action Refinement}
\label{algorithm}
\end{algorithm}




\subsection{Results and Discussion}

\begin{figure}[t!]
    \centering
    \includegraphics[width=\linewidth]{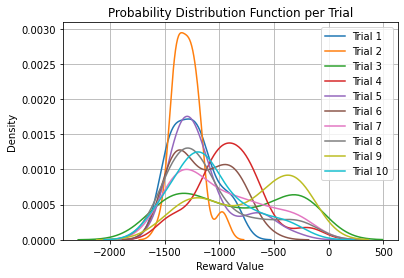}
    \caption{Probability Distribution Function per Trial. }
    \label{fig:GC}
\end{figure}

The results shown in the Fig.~\ref{fig:GC} illustrate the average reward value per trial on the model. Each represents the average and standard deviation regarding the reward values following every initialized point in space per trial. These trials are all uniform in that they all have performed 100 epochs, equating to 500 episodes in total to append to the experience deque continuously. Here each episode represents one sequence of states, actions, and rewards, which ends with a terminal state. For example, a person going from work to home is one episode. While repeating that cycle on a daily basis is considered multiple episodes. In this case, a terminal state of an episode is defined as an individual's ability to safely arrive at their destination on time. Additionally, one epoch equates to one step of action taken during the individual's journey. Furthermore, in our algorithm we determined the most accurate and cost-effective range to use was 100 epochs and 500 episodes. This is because the higher the epoch is the more data we have, which as a result will take longer to train the model. Thus, it becomes more expensive. The DDGP is a deep learning network, that can learn complicated patterns from very little amounts of data, hence we found 100 epochs and 500 episodes to be an excellent amount due to its fair amount of data. This chart is for observing the effectiveness of the trained model based on its states. While working with the DDPG model, it's essential to note that the initialization state is generally a random state in space that is selected, which is significant to how well your model performs. 

By observing the chart, both mean and standard deviation must be accounted for when determining which trial performed the best. An optimal policy yields a low standard deviation. Also, it has an average reward value that is high as possible.

\begin{figure}[t!]
    \centering
    \includegraphics[width=\linewidth]{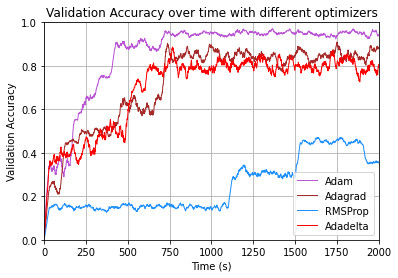}
    \caption{Reward Value per Episode. }
    \label{fig:LC}
\end{figure}

Respectfully, the different trials of the actor and the critic reward improvement based on their reward values as they iterate after every episode are in Fig.~\ref{fig:LC}. Also, we calculated the moving averages for each trial to account for future reward value in upcoming episodes. We elected to observe the moving averages and the future reward value to smoothen the data and eliminate the noise. Considering every episode is within a specific time frame, it would make sense to incorporate it with time series forecasting techniques such as moving averages.

Furthermore, Fig.~\ref{fig:GC} shows where most of the reward values following a given action are in the distribution. The DDPG model was trained and validated 10 times. After having trained the DDPG model 10 times, each model with randomly selected states, rewards, and actions from our original data, we added the distributions of the different reward values to find the model that resulted in the best reward value with the greatest number of times. This chart’s significance is presenting us with a clearer understanding of which model performed the best for our application. In the case of a real-world application using a DDPG model, it is strongly advised to use a model based on reward levels and the frequency of these higher reward values. This would verify that the application is functioning properly and would continue to develop, providing that a method is in place to train the model regularly.

The significance of Fig.~\ref{fig:LC} shows the difference in training time for the DDPG model based on different optimizers. These results are used to show the best optimizer for the implementation of our model. Regarding industry use, it would make sense to select an optimizer that will lead to the most accurate (in our case, the greatest reward after an action is taken by an actor) results, while also constraining the time necessary to fully train the model. Similarly, to the finding in section IV subsection C regarding training, reducing training time not only saves money and resource availability but also is a prerequisite in many real-world applications. Such real-world applications may include the field of robotics; in particular, the area of self-driving cars. This application would require lots of continuous data collection which may or may not be useful in getting predictions from the current running model; therefore, an application like this would require models to be updated frequently, which is possible when your models train faster.

\section{Conclusion}\label{conc}

In conclusion, we proposed a consensus system that uses the Raft algorithm paired with private blockchains for sharing resources in a MEC system. The network architecture is a cluster of edge nodes via a Hyperledger Fabric network. This ledger uses a private blockchain as its base. To test the effectiveness of our proposed design, we observed its learning rate and policy gradient using a DDPG model. We chose the policy gradient and validation accuracy of the system to measure the performance of our design. The training and testing yielded low standard deviations and maximized reward values. These results translate to an optimal policy gradient. Also, through the obtained validation accuracy, we observed continuous improvement within the learning rate of the model after each trial epoch. These results show that through the DDPG model, we have optimized our proposed system. Therefore, we present a design that improves the ability of edge-based distributed systems to provide more responsive and efficient services. 

\bibliographystyle{IEEEtran}
\bibliography{references}

\begin{thebibliography}{10}
\providecommand{\url}[1]{#1}
\csname url@samestyle\endcsname
\providecommand{\newblock}{\relax}
\providecommand{\bibinfo}[2]{#2}
\providecommand{\BIBentrySTDinterwordspacing}{\spaceskip=0pt\relax}
\providecommand{\BIBentryALTinterwordstretchfactor}{4}
\providecommand{\BIBentryALTinterwordspacing}{\spaceskip=\fontdimen2\font plus
\BIBentryALTinterwordstretchfactor\fontdimen3\font minus
  \fontdimen4\font\relax}
\providecommand{\BIBforeignlanguage}[2]{{%
\expandafter\ifx\csname l@#1\endcsname\relax
\typeout{** WARNING: IEEEtran.bst: No hyphenation pattern has been}%
\typeout{** loaded for the language `#1'. Using the pattern for}%
\typeout{** the default language instead.}%
\else
\language=\csname l@#1\endcsname
\fi
#2}}
\providecommand{\BIBdecl}{\relax}
\BIBdecl

\bibitem{9}
I.~Benyahia and M.~Hilali, ``An adaptive framework for distributed complex
  applications development,'' in \emph{Proceedings. 34th International
  Conference on Technology of Object-Oriented Languages and Systems - TOOLS
  34}, 2000, pp. 339--349.

\bibitem{b}
S.~Asad and A.~Refaey, ``On iot edge devices: Manifold unsupervised learning
  for som platforms,'' in \emph{2021 IEEE International Conference on Imaging
  Systems and Techniques (IST)}, 2021, pp. 1--5.

\bibitem{10}
Y.~Yun, Y.~Xia, B.~Behdani, and J.~C. Smith, ``Distributed algorithm for
  lifetime maximization in a delay-tolerant wireless sensor network with a
  mobile sink,'' \emph{IEEE Transactions on Mobile Computing}, vol.~12, no.~10,
  pp. 1920--1930, 2013.

\bibitem{f}
E.~Figetakis and A.~Refaey, ``Uav path planning using on-board ultrasound
  transducer arrays and edge support,'' in \emph{2021 IEEE International
  Conference on Communications Workshops (ICC Workshops)}, 2021, pp. 1--6.

\bibitem{11}
D.-S. Zhu and T.~C.-T. Lin, ``Causes and effects of e-service quality for
  online banking,'' in \emph{2012 13th ACIS International Conference on
  Software Engineering, Artificial Intelligence, Networking and
  Parallel/Distributed Computing}, 2012, pp. 357--361.

\bibitem{12}
X.~Wu and R.~Gao, ``The design and analysis of high performance online game
  server concurrent architecture,'' in \emph{2012 International Conference on
  Computer Science and Service System}, 2012, pp. 1662--1665.

\bibitem{g}
J.~Singh, A.~Refaey, and J.~Koilpillai, ``Adoption of the software-defined
  perimeter (sdp) architecture for infrastructure as a service,''
  \emph{Canadian Journal of Electrical and Computer Engineering}, vol.~43,
  no.~4, pp. 357--363, 2020.

\bibitem{h}
A.~Refaey, K.~Loukhaoukha, and A.~Dahmane, ``Cryptanalysis of stream cipher
  using density evolution,'' in \emph{2017 IEEE Conference on Communications
  and Network Security (CNS)}, 2017, pp. 382--383.

\bibitem{d}
A.~Vera-Rivera, A.~Refaey, and E.~Hossain, ``Task sharing and scheduling for
  edge computing servers using hyperledger fabric blockchain,'' in \emph{2021
  IEEE Globecom Workshops (GC Wkshps)}, 2021, pp. 1--6.

\bibitem{e}
A.~Vera-Rivera, E.~Hossain, and A.~R. Hussein, ``Exploring the intersection of
  consortium blockchain technologies and multi-access edge computing:
  Chronicles of a proof of concept demo,'' \emph{IEEE Open Journal of the
  Communications Society}, vol.~3, pp. 2203--2236, 2022.

\bibitem{4}
A.~V. Rivera, A.~Refaey, and E.~Hossain, ``A blockchain framework for secure
  task sharing in multi-access edge computing,'' 2020.

\bibitem{13}
P.~K. Kaushal, A.~Bagga, and R.~Sobti, ``Evolution of bitcoin and security risk
  in bitcoin wallets,'' in \emph{2017 International Conference on Computer,
  Communications and Electronics (Comptelix)}, 2017, pp. 172--177.

\bibitem{i}
A.~Refaey, R.~Niati, X.~Wang, and J.~Yves-Chouinard, ``Blind detection approach
  for ldpc, convolutional, and turbo codes in non-noisy environment,'' in
  \emph{2014 IEEE Conference on Communications and Network Security}, 2014, pp.
  502--503.

\bibitem{j}
A.~{Refaey}, W.~{Hou}, and L.~{Loukhaoukha}, ``Multilayer authentication for
  communication systems based on physical-layer attributes,'' \emph{Journal of
  Computer and Communications}, vol.~2, pp. 64--75, 2014.

\bibitem{k}
A.~{Refaey}, A.~{Sallam}, and A.~{Shami}, ``{On IoT applications: a proposed
  SDP framework for MQTT},'' \emph{Electronics Letters}, vol.~55, no.~22, pp.
  1201--1203, Oct. 2019.

\bibitem{l}
A.~Refaey, K.~Hammad, S.~Magierowski, and E.~Hossain, ``A blockchain policy and
  charging control framework for roaming in cellular networks,'' \emph{IEEE
  Network}, vol.~34, no.~3, pp. 170--177, 2020.

\bibitem{m}
A.~Moubayed, A.~Refaey, and A.~Shami, ``Software-defined perimeter (sdp): State
  of the art secure solution for modern networks,'' \emph{IEEE Network},
  vol.~33, no.~5, pp. 226--233, 2019.

\bibitem{mm}
W.-C. Shi and J.-P. Li, ``Research on consistency of distributed system based
  on paxos algorithm,'' in \emph{2012 International Conference on Wavelet
  Active Media Technology and Information Processing (ICWAMTIP)}, 2012, pp.
  257--259.

\bibitem{n}
P.~Kumar, A.~Moubayed, A.~Refaey, A.~Shami, and J.~Koilpillai, ``Performance
  analysis of sdp for secure internal enterprises,'' in \emph{2019 IEEE
  Wireless Communications and Networking Conference (WCNC)}, 2019, pp. 1--6.

\bibitem{3}
\BIBentryALTinterwordspacing
C.~Dwork, N.~Lynch, and L.~Stockmeyer, ``Consensus in the presence of partial
  synchrony,'' \emph{J. ACM}, vol.~35, no.~2, p. 288–323, Apr. 1988.
  [Online]. Available: \url{https://doi.org/10.1145/42282.42283}
\BIBentrySTDinterwordspacing

\bibitem{o}
A.~Sallam, A.~Refaey, and A.~Shami, ``Securing smart home networks with
  software-defined perimeter,'' in \emph{2019 15th International Wireless
  Communications \& Mobile Computing Conference (IWCMC)}, 2019, pp. 1989--1993.

\bibitem{nn}
S.~Wang, L.~Ouyang, Y.~Yuan, X.~Ni, X.~Han, and F.-Y. Wang,
  ``Blockchain-enabled smart contracts: Architecture, applications, and future
  trends,'' \emph{IEEE Transactions on Systems, Man, and Cybernetics: Systems},
  vol.~49, no.~11, pp. 2266--2277, 2019.

\bibitem{oo}
S.~N.~G. Gourisetti, M.~Mylrea, and H.~Patangia, ``Evaluation and demonstration
  of blockchain applicability framework,'' \emph{IEEE Transactions on
  Engineering Management}, vol.~67, no.~4, pp. 1142--1156, 2020.

\bibitem{2}
\BIBentryALTinterwordspacing
D.~Ongaro and J.~Ousterhout, ``In search of an understandable consensus
  algorithm,'' in \emph{2014 {USENIX} Annual Technical Conference ({USENIX}
  {ATC} 14)}.\hskip 1em plus 0.5em minus 0.4em\relax Philadelphia, PA: {USENIX}
  Association, Jun. 2014, pp. 305--319. [Online]. Available:
  \url{https://www.usenix.org/conference/atc14/technical-sessions/presentation/ongaro}
\BIBentrySTDinterwordspacing

\bibitem{p}
P.~Mirdita, Z.~Khaliq, A.~R. Hussein, and X.~Wang, ``Localization for
  intelligent systems using unsupervised learning and prediction approaches,''
  \emph{IEEE Canadian Journal of Electrical and Computer Engineering}, vol.~44,
  no.~4, pp. 443--455, 2021.

\bibitem{pp}
Z.~Bao, Q.~Wang, W.~Shi, L.~Wang, H.~Lei, and B.~Chen, ``When blockchain meets
  sgx: An overview, challenges, and open issues,'' \emph{IEEE Access}, vol.~8,
  pp. 170\,404--170\,420, 2020.

\bibitem{6}
K.~Croman, C.~Decker, I.~Eyal, A.~E. Gencer, A.~Juels, A.~Kosba, A.~Miller,
  P.~Saxena, E.~Shi, E.~G. Sirer, D.~Song, and R.~Wattenhofer, ``On scaling
  decentralized blockchains - (a position paper),'' in \emph{Financial
  Cryptography Workshops}, 2016.

\bibitem{7}
M.~Liu, F.~R. Yu, Y.~Teng, V.~C.~M. Leung, and M.~Song, ``Computation
  offloading and content caching in wireless blockchain networks with mobile
  edge computing,'' \emph{IEEE Transactions on Vehicular Technology}, vol.~67,
  no.~11, pp. 11\,008--11\,021, 2018.

\bibitem{q}
C.~DeSantis and A.~R. Hussein, ``Ai soc-based accelerator for speech
  classification ai soc-based accelerator for speech classification,''
  \emph{IEEE Canadian Journal of Electrical and Computer Engineering}, vol.~45,
  no.~3, pp. 222--231, 2022.

\bibitem{8}
H.~Lu, X.~Xu, K.~Zheng, and X.~Wang, ``An intelligent transaction migration
  scheme for raft-based private blockchain in internet of things
  applications,'' \emph{IEEE Communications Letters}, vol.~PP, pp. 1--1, 05
  2021.

\bibitem{1}
D.~{Huang}, X.~{Ma}, and S.~{Zhang}, ``Performance analysis of the raft
  consensus algorithm for private blockchains,'' \emph{IEEE Transactions on
  Systems, Man, and Cybernetics: Systems}, vol.~50, no.~1, pp. 172--181, 2020.

\bibitem{qq}
K.~Yue, Y.~Zhang, Y.~Chen, Y.~Li, L.~Zhao, C.~Rong, and L.~Chen, ``A survey of
  decentralizing applications via blockchain: The 5g and beyond perspective,''
  \emph{IEEE Communications Surveys \& Tutorials}, vol.~23, no.~4, pp.
  2191--2217, 2021.

\bibitem{a}
E.~Figetakis, A.~R. Hussein, and M.~Ulema, ``Evolved prevention strategies for
  6g networks through stochastic games and reinforcement learning,'' \emph{IEEE
  Networking Letters}, vol.~5, no.~3, pp. 164--168, 2023.

\bibitem{c}
E.~Figetakis and A.~Refaey, ``Autonomous mec selection in federated next-gen
  networks via deep reinforcement learning,'' in \emph{GLOBECOM 2023 - 2023
  IEEE Global Communications Conference}, 2023, pp. 2045--2050.

\end{thebibliography}

\end{document}